\RequirePackage{lineno}
\documentclass[prd,twocolumn,showpacs,superscriptaddress,floatfix]{revtex4}
\def\LineNumbers{false}
\def\AuthorList{true}
\def\PrlFigSpacing{true}

\usepackage{ifthen}
\usepackage{lineno}
\usepackage{epsfig}
\usepackage{amsmath}
\usepackage{comment}
\usepackage{longtable}
\usepackage{subfigure}
\usepackage{color}
\usepackage{xspace}
\usepackage{slashbox}

\usepackage{multirow}
\usepackage{array} 
\usepackage{booktabs}
\usepackage{dcolumn}
\usepackage[abs]{overpic}

\begin{document}

\newcommand{\vmet}{\ensuremath{\mbox{$\protect \raisebox{0.3ex}{$\not$}\vec{E}_T$}}\xspace}
\newcommand{\met} {\ensuremath{\mbox{$\protect \raisebox{0.3ex}{$\not$}E_T$}}\xspace}
\newcommand{\mht} {\ensuremath{\mbox{$\protect \raisebox{0.3ex}{$\not$}H_T$}}\xspace}
\newcommand{\vmpt}{\ensuremath{\mbox{$\protect \raisebox{0.3ex}{$\not$}\vec{p}_T$}}\xspace}
\newcommand{\mpt} {\ensuremath{\mbox{$\protect \raisebox{0.3ex}{$\not$}p_T$}}\xspace}

\newcommand{\vv}{\ensuremath{\nu\bar{\nu}}\xspace}
\newcommand{\lv}{\ensuremath{\ell\nu}\xspace}
\renewcommand{\ll}{\ensuremath{\ell\ell}\xspace}
\newcommand{\bb}{\ensuremath{b\bar{b}}\xspace}
\renewcommand{\tt}{\ensuremath{t\bar{t}}\xspace}
\newcommand{\metbb}{\ensuremath{\met+\bb}\xspace}

\newcommand{\nnjer}{\ensuremath{\mathrm{NN}_\mathrm{JER}}\xspace}
\newcommand{\nnqcd}{\ensuremath{\mathrm{NN}_\mathrm{QCD}}\xspace}
\newcommand{\nnsig}{\ensuremath{\mathrm{NN}_\mathrm{SIG}}\xspace}

\newcommand{\vh}{\ensuremath{V\!H}\xspace}
\newcommand{\wh}{\ensuremath{W\!H}\xspace}
\newcommand{\zh}{\ensuremath{Z\!H}\xspace}

\newcommand{\hbb}{\ensuremath{H \to \bb}\xspace}
\newcommand{\invfb}{fb$^{-1}$\xspace}
\newcommand{\gevcc}{\ensuremath{\mathrm{GeV}/c^2}\xspace} 
\newcommand{\gevc}{GeV$/c$\xspace} 
\newcommand{\gev}{GeV\xspace} 
\newcommand{\tev}{TeV\xspace} 

\renewcommand{\b}{\ensuremath{b}\xspace}

\ifthenelse{\equal{\LineNumbers}{true}}{
 \linenumbers
}

\title{Updated search for the standard model Higgs boson in events with jets and missing transverse energy using the full CDF data set}

\ifthenelse{\equal{\AuthorList}{true}}{
  \affiliation{Institute of Physics, Academia Sinica, Taipei, Taiwan 11529, Republic of China}
\affiliation{Argonne National Laboratory, Argonne, Illinois 60439, USA}
\affiliation{University of Athens, 157 71 Athens, Greece}
\affiliation{Institut de Fisica d'Altes Energies, ICREA, Universitat Autonoma de Barcelona, E-08193, Bellaterra (Barcelona), Spain}
\affiliation{Baylor University, Waco, Texas 76798, USA}
\affiliation{Istituto Nazionale di Fisica Nucleare Bologna, $^{ee}$University of Bologna, I-40127 Bologna, Italy}
\affiliation{University of California, Davis, Davis, California 95616, USA}
\affiliation{University of California, Los Angeles, Los Angeles, California 90024, USA}
\affiliation{Instituto de Fisica de Cantabria, CSIC-University of Cantabria, 39005 Santander, Spain}
\affiliation{Carnegie Mellon University, Pittsburgh, Pennsylvania 15213, USA}
\affiliation{Enrico Fermi Institute, University of Chicago, Chicago, Illinois 60637, USA}
\affiliation{Comenius University, 842 48 Bratislava, Slovakia; Institute of Experimental Physics, 040 01 Kosice, Slovakia}
\affiliation{Joint Institute for Nuclear Research, RU-141980 Dubna, Russia}
\affiliation{Duke University, Durham, North Carolina 27708, USA}
\affiliation{Fermi National Accelerator Laboratory, Batavia, Illinois 60510, USA}
\affiliation{University of Florida, Gainesville, Florida 32611, USA}
\affiliation{Laboratori Nazionali di Frascati, Istituto Nazionale di Fisica Nucleare, I-00044 Frascati, Italy}
\affiliation{University of Geneva, CH-1211 Geneva 4, Switzerland}
\affiliation{Glasgow University, Glasgow G12 8QQ, United Kingdom}
\affiliation{Harvard University, Cambridge, Massachusetts 02138, USA}
\affiliation{Division of High Energy Physics, Department of Physics, University of Helsinki and Helsinki Institute of Physics, FIN-00014, Helsinki, Finland}
\affiliation{University of Illinois, Urbana, Illinois 61801, USA}
\affiliation{The Johns Hopkins University, Baltimore, Maryland 21218, USA}
\affiliation{Institut f\"{u}r Experimentelle Kernphysik, Karlsruhe Institute of Technology, D-76131 Karlsruhe, Germany}
\affiliation{Center for High Energy Physics: Kyungpook National University, Daegu 702-701, Korea; Seoul National University, Seoul 151-742, Korea; Sungkyunkwan University, Suwon 440-746, Korea; Korea Institute of Science and Technology Information, Daejeon 305-806, Korea; Chonnam National University, Gwangju 500-757, Korea; Chonbuk National University, Jeonju 561-756, Korea; Ewha Womans University, Seoul, 120-750, Korea}
\affiliation{Ernest Orlando Lawrence Berkeley National Laboratory, Berkeley, California 94720, USA}
\affiliation{University of Liverpool, Liverpool L69 7ZE, United Kingdom}
\affiliation{University College London, London WC1E 6BT, United Kingdom}
\affiliation{Centro de Investigaciones Energeticas Medioambientales y Tecnologicas, E-28040 Madrid, Spain}
\affiliation{Massachusetts Institute of Technology, Cambridge, Massachusetts 02139, USA}
\affiliation{Institute of Particle Physics: McGill University, Montr\'{e}al, Qu\'{e}bec H3A~2T8, Canada; Simon Fraser University, Burnaby, British Columbia V5A~1S6, Canada; University of Toronto, Toronto, Ontario M5S~1A7, Canada; and TRIUMF, Vancouver, British Columbia V6T~2A3, Canada}
\affiliation{University of Michigan, Ann Arbor, Michigan 48109, USA}
\affiliation{Michigan State University, East Lansing, Michigan 48824, USA}
\affiliation{Institution for Theoretical and Experimental Physics, ITEP, Moscow 117259, Russia}
\affiliation{University of New Mexico, Albuquerque, New Mexico 87131, USA}
\affiliation{The Ohio State University, Columbus, Ohio 43210, USA}
\affiliation{Okayama University, Okayama 700-8530, Japan}
\affiliation{Osaka City University, Osaka 588, Japan}
\affiliation{University of Oxford, Oxford OX1 3RH, United Kingdom}
\affiliation{Istituto Nazionale di Fisica Nucleare, Sezione di Padova-Trento, $^{ff}$University of Padova, I-35131 Padova, Italy}
\affiliation{University of Pennsylvania, Philadelphia, Pennsylvania 19104, USA}
\affiliation{Istituto Nazionale di Fisica Nucleare Pisa, $^{gg}$University of Pisa, $^{hh}$University of Siena and $^{ii}$Scuola Normale Superiore, I-56127 Pisa, Italy, $^{mm}$INFN Pavia and University of Pavia, I-27100 Pavia, Italy}
\affiliation{University of Pittsburgh, Pittsburgh, Pennsylvania 15260, USA}
\affiliation{Purdue University, West Lafayette, Indiana 47907, USA}
\affiliation{University of Rochester, Rochester, New York 14627, USA}
\affiliation{The Rockefeller University, New York, New York 10065, USA}
\affiliation{Istituto Nazionale di Fisica Nucleare, Sezione di Roma 1, $^{jj}$Sapienza Universit\`{a} di Roma, I-00185 Roma, Italy}
\affiliation{Texas A\&M University, College Station, Texas 77843, USA}
\affiliation{Istituto Nazionale di Fisica Nucleare Trieste/Udine; $^{nn}$University of Trieste, I-34127 Trieste, Italy; $^{kk}$University of Udine, I-33100 Udine, Italy}
\affiliation{University of Tsukuba, Tsukuba, Ibaraki 305, Japan}
\affiliation{Tufts University, Medford, Massachusetts 02155, USA}
\affiliation{University of Virginia, Charlottesville, Virginia 22906, USA}
\affiliation{Waseda University, Tokyo 169, Japan}
\affiliation{Wayne State University, Detroit, Michigan 48201, USA}
\affiliation{University of Wisconsin, Madison, Wisconsin 53706, USA}
\affiliation{Yale University, New Haven, Connecticut 06520, USA}

\author{T.~Aaltonen}
\affiliation{Division of High Energy Physics, Department of Physics, University of Helsinki and Helsinki Institute of Physics, FIN-00014, Helsinki, Finland}
\author{S.~Amerio}
\affiliation{Istituto Nazionale di Fisica Nucleare, Sezione di Padova-Trento, $^{ff}$University of Padova, I-35131 Padova, Italy}
\author{D.~Amidei}
\affiliation{University of Michigan, Ann Arbor, Michigan 48109, USA}
\author{A.~Anastassov$^x$}
\affiliation{Fermi National Accelerator Laboratory, Batavia, Illinois 60510, USA}
\author{A.~Annovi}
\affiliation{Laboratori Nazionali di Frascati, Istituto Nazionale di Fisica Nucleare, I-00044 Frascati, Italy}
\author{J.~Antos}
\affiliation{Comenius University, 842 48 Bratislava, Slovakia; Institute of Experimental Physics, 040 01 Kosice, Slovakia}
\author{G.~Apollinari}
\affiliation{Fermi National Accelerator Laboratory, Batavia, Illinois 60510, USA}
\author{J.A.~Appel}
\affiliation{Fermi National Accelerator Laboratory, Batavia, Illinois 60510, USA}
\author{T.~Arisawa}
\affiliation{Waseda University, Tokyo 169, Japan}
\author{A.~Artikov}
\affiliation{Joint Institute for Nuclear Research, RU-141980 Dubna, Russia}
% ...

\collaboration{CDF Collaboration\footnote{With visitors from
$^a$Istituto Nazionale di Fisica Nucleare, Sezione di Cagliari, 09042 Monserrato (Cagliari), Italy,
$^b$University of California Irvine, Irvine, CA 92697, USA,
$^c$University of California Santa Barbara, Santa Barbara, CA 93106, USA,
$^d$University of California Santa Cruz, Santa Cruz, CA 95064, USA,
$^e$Institute of Physics, Academy of Sciences of the Czech Republic, 182~21, Czech Republic,
$^f$CERN, CH-1211 Geneva, Switzerland,
$^g$Cornell University, Ithaca, NY 14853, USA,
$^h$University of Cyprus, Nicosia CY-1678, Cyprus,
$^i$Office of Science, U.S. Department of Energy, Washington, DC 20585, USA,
$^j$University College Dublin, Dublin 4, Ireland,
$^k$ETH, 8092 Z\"{u}rich, Switzerland,
$^l$University of Fukui, Fukui City, Fukui Prefecture, Japan 910-0017,
$^m$Universidad Iberoamericana, Lomas de Santa Fe, M\'{e}xico, C.P. 01219, Distrito Federal,
$^n$University of Iowa, Iowa City, IA 52242, USA,
$^o$Kinki University, Higashi-Osaka City, Japan 577-8502,
$^p$Kansas State University, Manhattan, KS 66506, USA,
$^q$Brookhaven National Laboratory, Upton, NY 11973, USA,
$^r$University of Manchester, Manchester M13 9PL, United Kingdom,
$^s$Queen Mary, University of London, London, E1 4NS, United Kingdom,
$^t$University of Melbourne, Victoria 3010, Australia,
$^u$Muons, Inc., Batavia, IL 60510, USA,
$^v$Nagasaki Institute of Applied Science, Nagasaki 851-0193, Japan,
$^w$National Research Nuclear University, Moscow 115409, Russia,
$^x$Northwestern University, Evanston, IL 60208, USA,
$^y$University of Notre Dame, Notre Dame, IN 46556, USA,
$^z$Universidad de Oviedo, E-33007 Oviedo, Spain,
$^{aa}$CNRS-IN2P3, Paris, F-75205 France,
$^{bb}$Texas Tech University, Lubbock, TX 79609, USA,
$^{cc}$Universidad Tecnica Federico Santa Maria, 110v Valparaiso, Chile,
$^{dd}$Yarmouk University, Irbid 211-63, Jordan,
$^{ll}$Universite catholique de Louvain, 1348 Louvain-La-Neuve, Belgium,
$^{oo}$University of Z\"{u}rich, 8006 Z\"{u}rich, Switzerland,
$^{pp}$Massachusetts General Hospital and Harvard Medical School, Boston, MA 02114 USA,
$^{qq}$Hampton University, Hampton, VA 23668, USA,
$^{rr}$Los Alamos National Laboratory, Los Alamos, NM 87544, USA
}}
\noaffiliation
 
}

\ifthenelse{\equal{\AuthorList}{false}}{
\begin{figure}  
  \leftline{\includegraphics[scale=0.5]{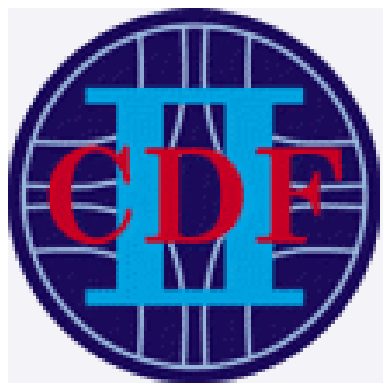}\hfill
    CDF/PUB/EXOTIC/CDFR/10903}
\end{figure}

\vspace*{0.5cm} 

\author{The CDF Collaboration}
\affiliation{URL http://www-cdf.fnal.gov}
}

\date{\today}

\begin{abstract}
We present an updated search for the Higgs boson produced in
association with a vector boson in the final state with missing
transverse energy and two jets.  We use the full CDF data set
corresponding to an integrated luminosity of 9.45 \invfb at a
proton-antiproton center-of-mass energy of $\sqrt{s}=1.96$ \tev.  New
to this analysis is the inclusion of a $b$-jet identification
algorithm specifically optimized for \hbb searches.  Across the Higgs
boson mass range $90 \le m_H \le 150$ \gevcc, the expected 95\%
credibility level upper limits on the $\vh$ production cross section
times the \hbb branching fraction are improved by an average of 14\%
relative to the previous analysis. At a Higgs boson mass of $125$
\gevcc, the observed (expected) limit is 3.06 (3.33) times the
standard model prediction, corresponding to one of the most sensitive
searches to date in this final state.
\end{abstract}



\maketitle

\section{Introduction}

In the standard model of particle physics (SM)~\cite{sm}, the
mechanism of electroweak symmetry breaking generates a massive scalar
boson called the Higgs boson ($H$)~\cite{higgs}.  Over the last few
decades there has been an intensive effort to uncover experimental
evidence of the existence of the Higgs boson.  Recently, the CMS and
ATLAS collaborations reported the observation of a new boson with a
mass of approximately 125
\gevcc~\cite{higgs_discovery_accepted}. While the production and decay
of this particle are consistent with expectations for the SM Higgs
boson, many of its properties have yet to be established.  In
particular, the relative coupling strengths of this boson to quarks,
leptons, and other bosons are important in understanding whether it is
the SM Higgs boson or another state.  While the sensitivities of the
CMS and ATLAS analyses were primarily influenced by decays of this
particle into $Z$ bosons, $W$ bosons, and photons, the sensitivity of
the low-mass Higgs boson analyses of the CDF and D0 collaborations is
largely from decays to pairs of $b$ quarks.  Recent results from CDF
and D0 show evidence of an excess of events consistent with a 125
\gevcc SM Higgs boson decaying to $b$ quarks~\cite{hbb_tev}. 
However, it is not yet known if this excess can be attributed to the
same particle observed by the ATLAS and CMS collaborations and further
investigation is warranted.

In the SM, the dominant decay channel for a low-mass Higgs boson ($m_H
\le 135$ \gevcc) is to the \bb final state.  At the Tevatron, pairs
of $b$ quarks are produced via the strong interaction (``QCD multijet''
background) with a cross section much larger than that predicted for
Higgs boson production followed by \hbb decay.  Searching for direct
Higgs boson production is, therefore, very difficult and far less
sensitive than searching for it in processes where the SM Higgs boson
is produced in association with a weak vector boson $V$ (where $V$
represents the $W$ or $Z$ boson).  The leptonic decay of the vector
boson provides a distinct signature, enabling significant suppression
of QCD multijet events. Furthermore, selecting events in which jets
are identified as being consistent with the fragmentation of $b$
quarks (``$b$ tagging'') additionally improves the
signal-to-background ratio in low-mass SM Higgs boson searches.

One of the most sensitive SM Higgs boson search channels at the
Tevatron is the $\vh\to\metbb$ final state, where \met represents
the missing tranverse energy resulting from neutrinos or unidentified
charged leptons in the event.  This article reports an update to the
previous CDF analysis in the \metbb search channel~\cite{metbb}; the
same data are analyzed, but the \b-tagging strategy is significantly
improved.  The complete \metbb analysis method has been described
previously~\cite{metbb} and will only be briefly reviewed.  The data
correspond to an integrated luminosity of $9.45$ \invfb, collected in
proton-antiproton collisions at a center-of-mass energy of $\sqrt{s} =
1.96$ \tev.

\section{\label{sec:evsel} CDF Detector and Event Selection}

The CDF~II detector is described in detail elsewhere~\cite{cdf,
geometry}.  It features a cylindrical silicon detector and drift wire
tracking system inside a superconducting solenoid, surrounded by
projective calorimeters and muon detectors.  Calorimeter energy
deposits are clustered into jets using a cone algorithm with an
opening angle of $\Delta
R\equiv\sqrt{(\Delta\phi)^2+(\Delta\eta)^2}=0.4$~\cite{Bhatti_2005ai}.
High-$p_T$ electron candidates are identified by matching
charged-particle tracks in the inner tracking
systems~\cite{silicon,cot} with energy deposits in the electromagnetic
calorimeters~\cite{em}.  Muon candidates are identified by matching
tracks with muon-detector track segments~\cite{muons}.  The
hermeticity of the calorimeter in the pseudorapidity range $|\eta| <
2.4$ provides reliable reconstruction of the missing transverse
energy~\cite{MET,had}.


Events are selected during online data taking if they contain either
$\met(\mbox{cal}) > 45$ \gev, or $\met(\mbox{cal}) > 35$ \gev and at
least two jets.  In the analysis, we further require that events
contain no identified electron or muon, and $\met > 35$ \gev after
corrections for instrumental effects in jet reconstruction are
applied~\cite{Bhatti_2005ai}.  The two jets of greatest $E_T$ in the
event are required to have transverse energies that satisfy $25 <
E_T^{j_1} < 200$ \gev and $20 < E_T^{j_2} < 120$ \gev, respectively,
according to a jet-energy determination based on calorimeter deposits
and track momentum measurements~\cite{h1}.  This selects candidate
events consistent with the $\zh\to\vv\bb$ process.  Because $\tau$
leptons are not explicitly reconstructed and some electrons and muons
escape detection or reconstruction, events from the $\wh\to\lv\bb$
process are also expected to contribute significantly.  To gain
sensitivity in events with an unidentified $\tau$ lepton, we therefore
also accept events where the third-most energetic jet satisfies $15 <
E_T^{j_3} < 100$ \gev.  We reject events with four reconstructed jets,
where each jet exceeds the minimum transverse energy threshold ($E_T >
15$ \gev) and has pseudorapidity $|\eta| < 2.4$.  To reduce
contamination from QCD multijet events that exhibit \met generated via
jet mismeasurement, the angles between the \vmet and the directions of
the second and (if present) third jets are required to be greater than
0.4 radians.  To ensure that both leading-$E_T$ jets are reconstructed
within the silicon detector acceptance, they are required to satisfy
$|\eta| < 2$, where at least one of them must satisfy $|\eta| < 0.9$.
The QCD multijet background is additionally reduced by 35\% using a
neural-network regression algorithm that incorporates electromagnetic-
and hadronic-calorimeter quantities to account for jet-energy
mismeasurements.  

\begin{table}[b]
  \setlength{\extrarowheight}{3pt}
\caption{\label{tbl:tag_effs} Comparison of \b-tagging efficiencies 
per signal event in the tag categories of this analysis and the
previous one~\cite{metbb}.  Jets tagged by the \textsc{secvtx}
\b-tagging algorithm are labeled ``S'', and those that are tagged by
the \textsc{jetprob} algorithm but not \textsc{secvtx} are labeled
``J''.  There is no overlap between the tag categories of a given
analysis by design.}
\begin{tabular*}{\linewidth}{@{\extracolsep{\fill}}lcc} \hline\hline
\multirow{2}{*}{Tag category}    & \multicolumn{2}{c}{\b-tagging efficiency per event} \\ \cline{2-3}
                                 &  Ref.~\cite{metbb}  & This analysis \\ \hline
Two tight $b$ tags               & 13.7\% (SS)         & 18.1\% (TT) \\ 
One tight and one loose $b$ tag  & 13.1\% (SJ)         & 14.6\% (TL) \\ 
Only one tight $b$ tag           & 31.4\% (1S)         & 31.6\% (1T) \\ \hline\hline
\end{tabular*}  
\end{table}

\section{\label{sec:btag} \boldmath $b$-jet Identification Algorithm}

This analysis employs a multivariate \b-tagging algorithm
(\textsc{hobit}) specifically optimized for \hbb
searches~\cite{hobit}.  The algorithm incorporates quantities from
various CDF \b-tagging algorithms as input variables, and it assigns
an output value $v$ to each jet based on the probability that the jet
originates from the fragmentation of a $b$ quark.  Jets initiated by
$b$ quarks tend to cluster at values close to 1, whereas those
initiated by light-flavor quarks are more likely to populate the
region near $-1$.  Two operating regions are used: jets with $v \ge
0.98$ are considered to be tightly tagged (T), whereas jets with $0.72
< v < 0.98$ are loosely tagged (L).  Analogous to the previous
analysis, we accept events assigned to one of three categories based
on the tag quality of the two leading-$E_T$ jets: both jets are
tightly tagged (TT); one jet is tightly tagged, and the other loosely
tagged (TL); and only one jet is tightly tagged (1T).  The tag
categories used in both analyses and the associated tagging
efficiencies of Higgs boson signal events are given in
Table~\ref{tbl:tag_effs}.  As can be seen, the \textsc{hobit}
algorithm achieves a 32\% (11\%) relative improvement in the tagging
efficiency of signal events into the double-tight (tight-loose)
category.  The \emph{preselection sample} consists of events that
satisfy all of the above selection criteria.

\section{QCD Multijet Background Model}

In the preselection sample, the dominant background to the Higgs boson
signal is still that of QCD multijet production. Other non-neglible
backgrounds are those from singly- and pair-produced top quarks
(``top''), $V$-plus-heavy-flavor jets, diboson production ($VV$), and
jets from electroweak processes that are incorrectly tagged as $b$
jets (``electroweak mistags'').  The modeling of each background is
described in Ref.~\cite{metbb}.  A QCD multijet background model is
derived by looking at data events in a control region where $\met <
70$ \gev and the angle between the \vmet and second jet is less than
0.4 radians.  The sample of events that satisfy these criteria
consists almost entirely of QCD multijet contributions.  For tag
category $i$ (where $i = $ 1T, TL, or TT), a multivariable probability
density function $f_i$ is formed by taking the ratio between tagged
and pretagged events as a function of several variables. Four of those
variables are the same as in Ref.~\cite{metbb}: the scalar sum of jet
transverse energies $H_T$, the missing track transverse momentum of
the event \mpt, and the charge fractions ($\sum_i p_{T}^i/E_{T}$,
where the sum is over the tracks within the jet cone) of the first-
and second-most energetic jets.  To improve the modeling of the QCD
multijet background, we include two more parameters in the probability
density function: the number of reconstructed vertices in the event,
which is correlated with the topological variables used in the
multivariate discriminants (see Sec.~\ref{sec:mvas}); and $p_\perp^\mu
= p_{\mu1}\sin(\hat{\boldsymbol{\mu}}_1,\hat{\mathbf{j}}_1) +
p_{\mu2}\sin(\hat{\boldsymbol{\mu}}_2,\hat{\mathbf{j}}_2)$, where
$p_{\mu i}$ represents the momentum of the most energetic muon (if one
exists) within the cone of jet $i$, and
$\sin(\hat{\boldsymbol{\mu}}_i,\hat{\mathbf{j}}_i)$ is the sine of the
angle between the muon and jet directions.  The $p_\perp^\mu$ variable
tends to be large for jets in which the initiating $b$ quark decays
semileptonically through $b \to c\ell\nu$.

\begin{figure}
\ifthenelse{\equal{\PrlFigSpacing}{true}}
{
  \begin{overpic}[width=\linewidth]{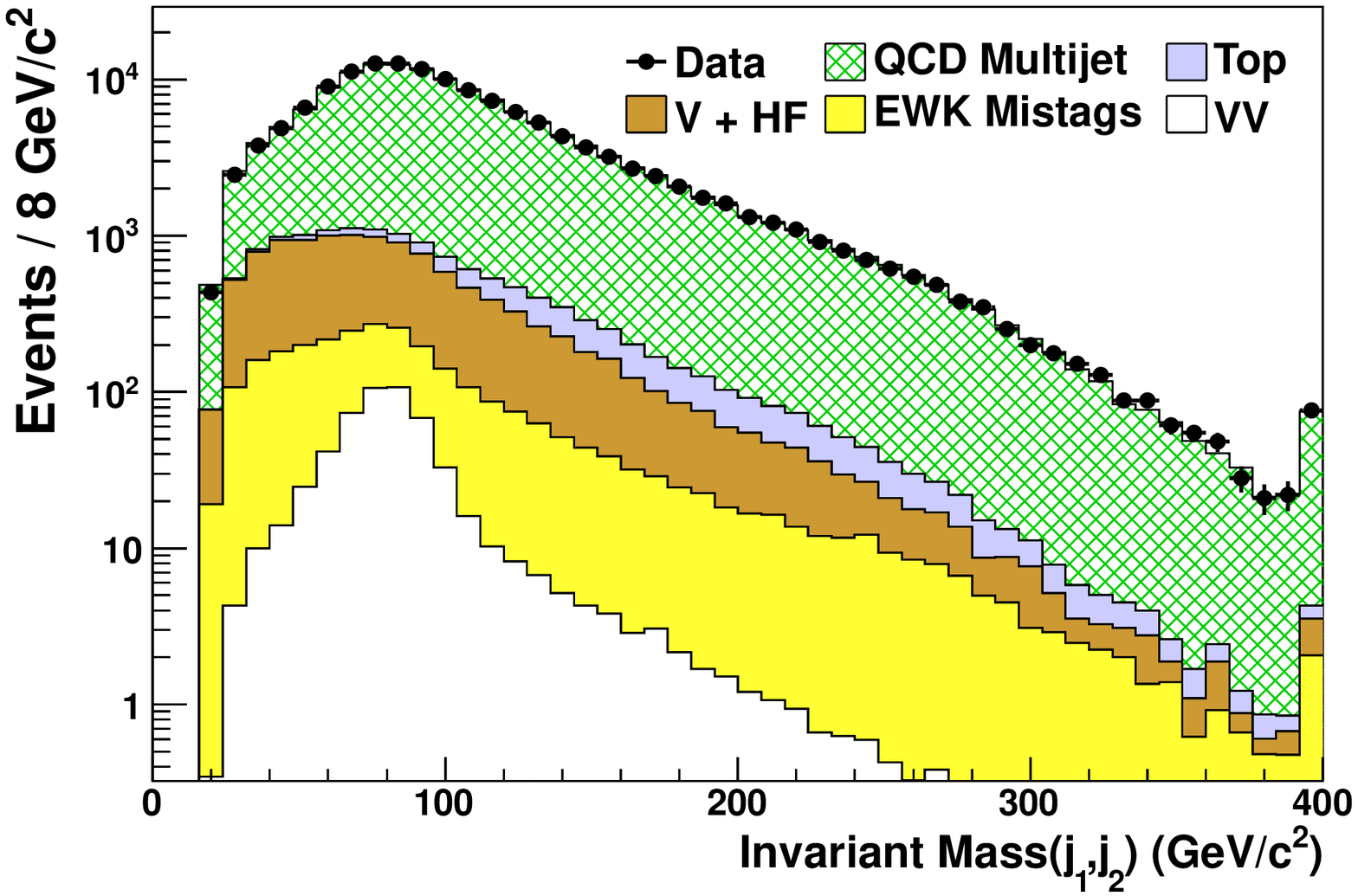}
  \put(217,120){\bf(a)}
  \end{overpic}
  \begin{overpic}[width=\linewidth]{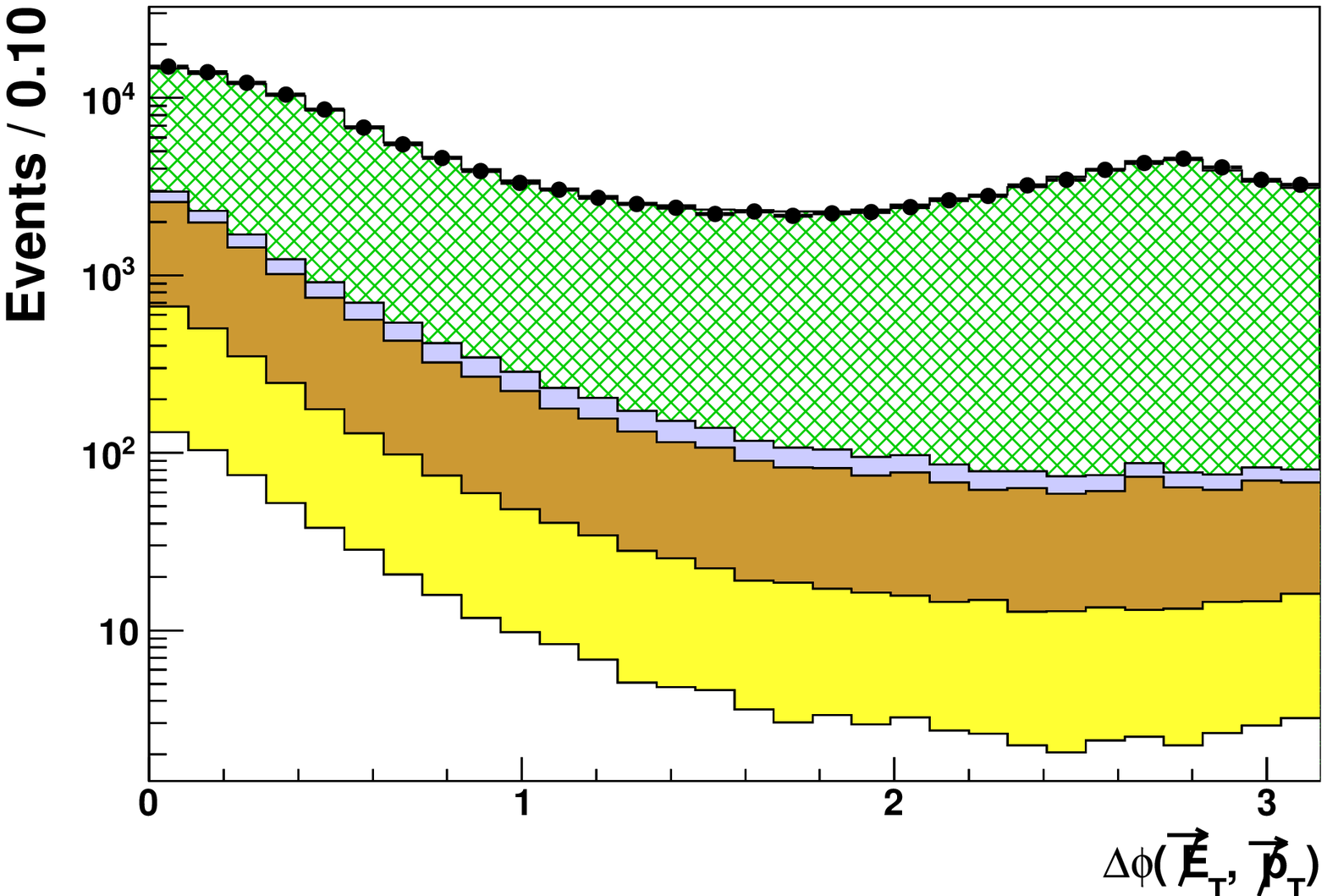}
  \put(217,138){\bf(b)}
  \end{overpic} 
  \begin{overpic}[width=\linewidth]{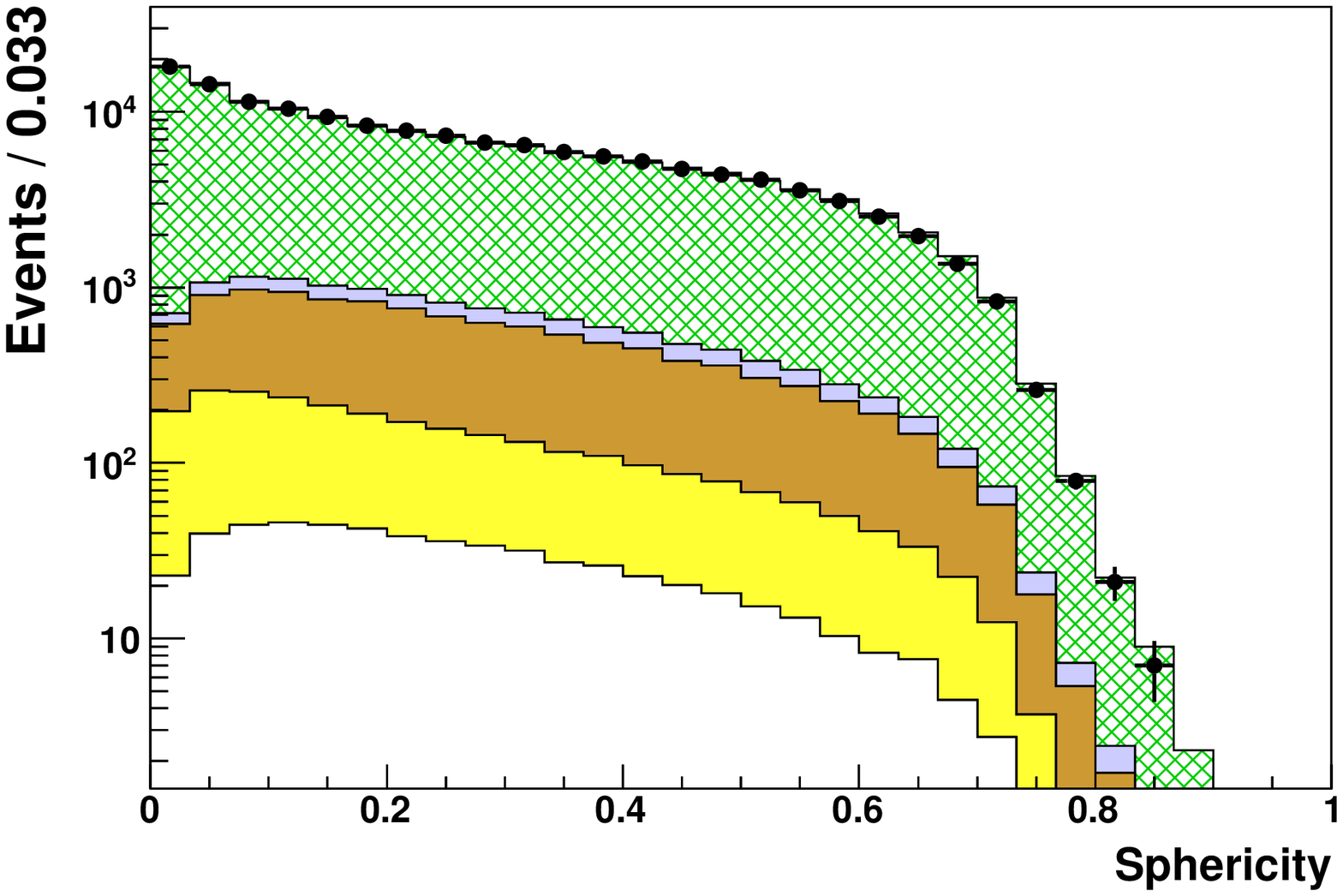}
  \put(217,130){\bf(c)}
  \end{overpic} 
}
{
  \begin{overpic}[width=0.6\linewidth]{figures/presel_MJJ_allTags_CENT_LOG.eps}
  \put(217,120){\bf(a)}
  \end{overpic}
  \begin{overpic}[width=0.6\linewidth]{figures/presel_DPhi_allTags_CENT_LOG.eps}
  \put(217,155){\bf(b)}
  \end{overpic} 
  \begin{overpic}[width=0.6\linewidth]{figures/presel_Sphericity_allTags_CENT_LOG.eps}
  \put(217,130){\bf(c)}
  \end{overpic} 
}
\caption{ Validation of the background model for all tagged events in the
preselection sample for (a) the invariant mass of the two leading
jets, (b) the angle between the \vmet and \vmpt, and (c) the
sphericity of the jets in the event.}
\label{fig:comps}
\end{figure}

A QCD multijet model is determined for each of the 1T, TL, and TT
categories by weighting the untagged data in the preselection sample
according to the $f_{\mathrm{1T}}$, $f_{\mathrm{TL}}$, and
$f_{\mathrm{TT}}$ probability density functions, respectively.  To
determine the appropriate normalization for a given category, the
tagged $VV$, top, $V$-plus-heavy-flavor, and electroweak mistag
background estimates are subtracted from the tagged data, and the
multijet prediction is scaled to that difference.  To validate the
background modeling, we compare tagged data and the corresponding
combined background prediction in multiple control regions~\cite{crs}
for various kinematic, angular, and event-shape variables, which are
included later on as inputs to multivariate discriminants that
separate signal and background processes.  Shown in
Fig.~\ref{fig:comps} are data-modeling comparisons of all tagged
events in the preselection sample for the invariant dijet mass
(kinematic), the angle between the \vmet and \vmpt directions
$\Delta\phi(\vmet,\vmpt)$ (angular), and jet sphericity (event
shape)~\cite{sphericity} variables.  The good agreement found in each
distribution is representative of all variables included in the
neural-network discriminants described below.

\section{\label{sec:mvas} Multivariate Discriminants}

To optimally separate Higgs boson signal from background, a staged
multivariate approach is used.  A first neural network
\nnqcd is trained to discriminate between QCD multijet and signal 
processes.  Events that satisfy a minimum \nnqcd threshold requirement
are subjected to a second neural network \nnsig, designed to separate
the signal from the remaining SM backgrounds.

\begin{figure}[t]
\ifthenelse{\equal{\PrlFigSpacing}{true}}
{
  \begin{overpic}[width=\linewidth]{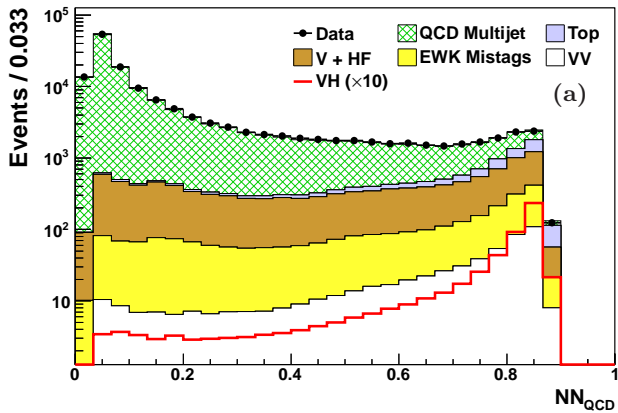} 
  \put(210,120){\bf(a)}
  \end{overpic}
}
{
  \begin{overpic}[width=0.6\linewidth]{figures/presel_NNQCD_allTags_CENT_LOG.eps} 
  \put(240,120){\bf(a)}
  \end{overpic}
}
\caption{\label{fig:nnqcd} 
The distribution of the \nnqcd discriminant for tagged data events in
the preselection sample in comparison with modeled background expectations.}
\end{figure}

The \nnqcd discriminant is trained using equal event yields of QCD
multijet-modeled background and \vh signal processes.  As in the
previous analysis, the collection of input variables to the \nnqcd
algorithm includes kinematic, angular, and event-shape
quantities~\cite{metbb,Potamianos:2011}, each of which is validated
with tagged data in the preselection sample.  Figure~\ref{fig:nnqcd}
shows the \nnqcd distribution for tagged events satisfying the
preselection criteria.  By imposing a minimum \nnqcd requirement of
0.6 (which defines the \emph{signal region}), 87\% of the signal is
retained while 90\% of the QCD multijet background is rejected.
Table~\ref{tbl:eventYields} shows the expected number of signal and
background events and the observed data events in the signal region.
For a Higgs boson mass of 125 \gevcc, we expect 19 signal events in
the 1T category and roughly 11 signal events in both the TL and TT
categories.

\begin{figure}
\ifthenelse{\equal{\PrlFigSpacing}{true}}
{
  \begin{overpic}[width=\linewidth]{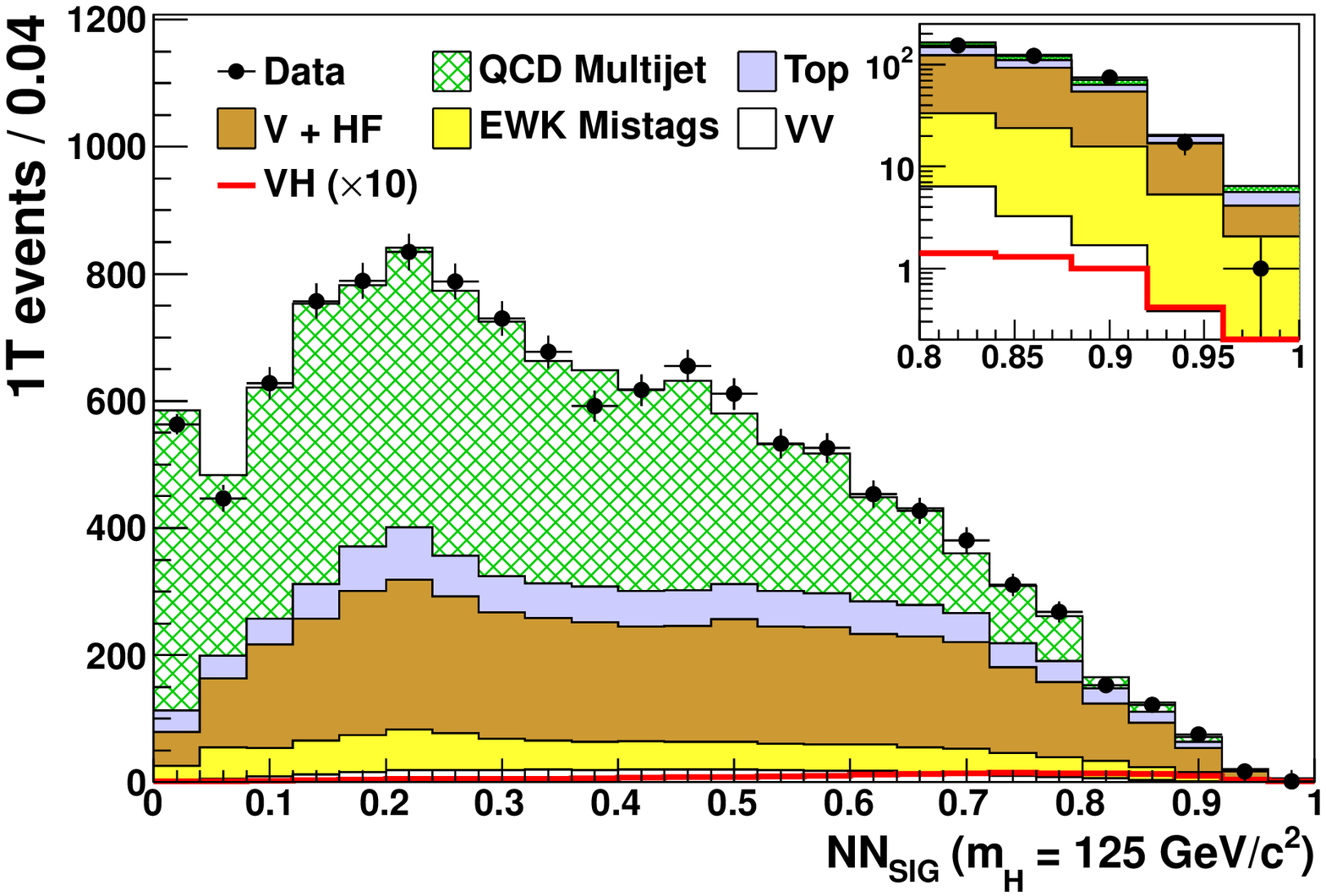}
  \put(210,70){\bf(a)}
  \end{overpic}
  \begin{overpic}[width=\linewidth]{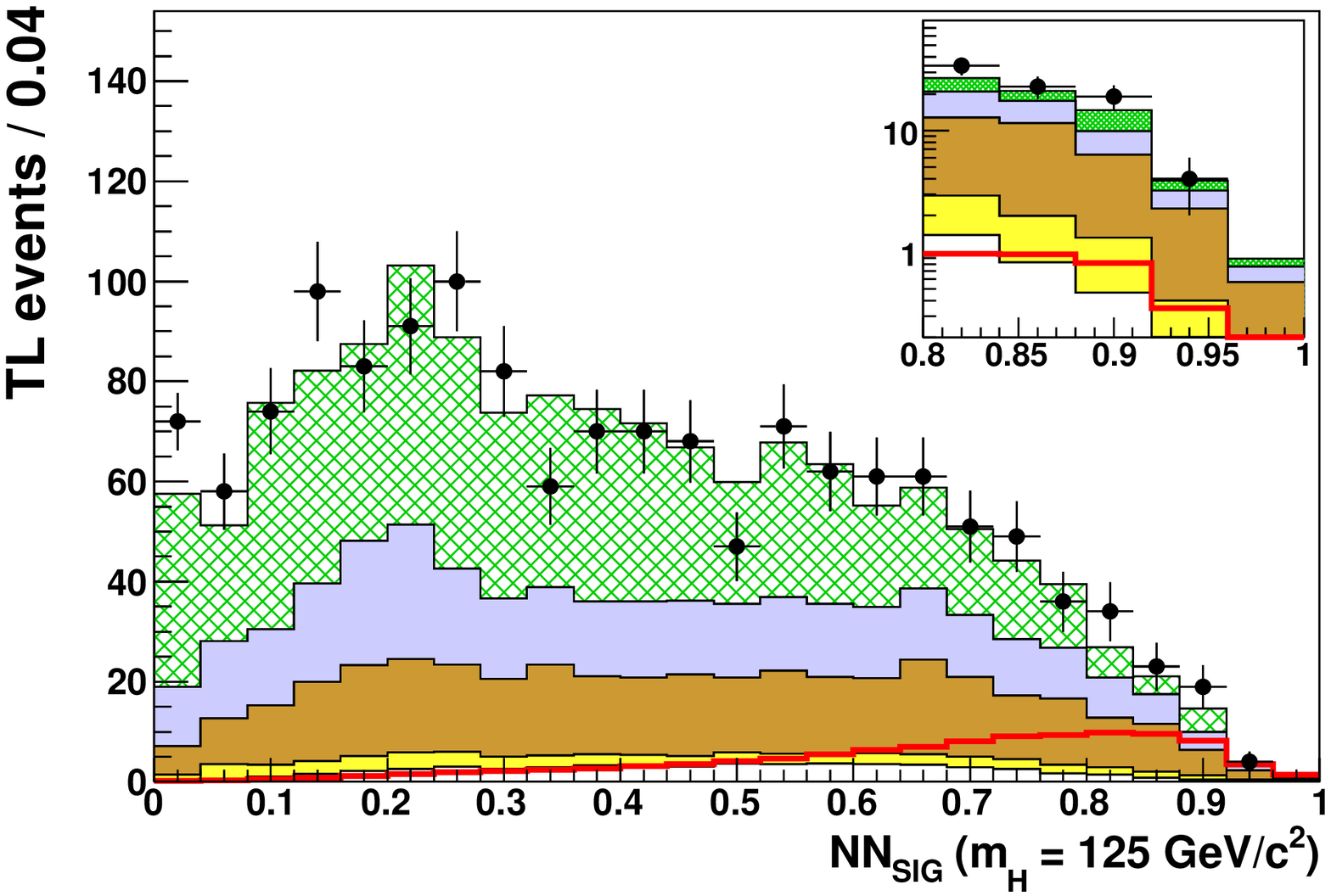}
  \put(210,70){\bf(b)}
  \end{overpic}
  \begin{overpic}[width=\linewidth]{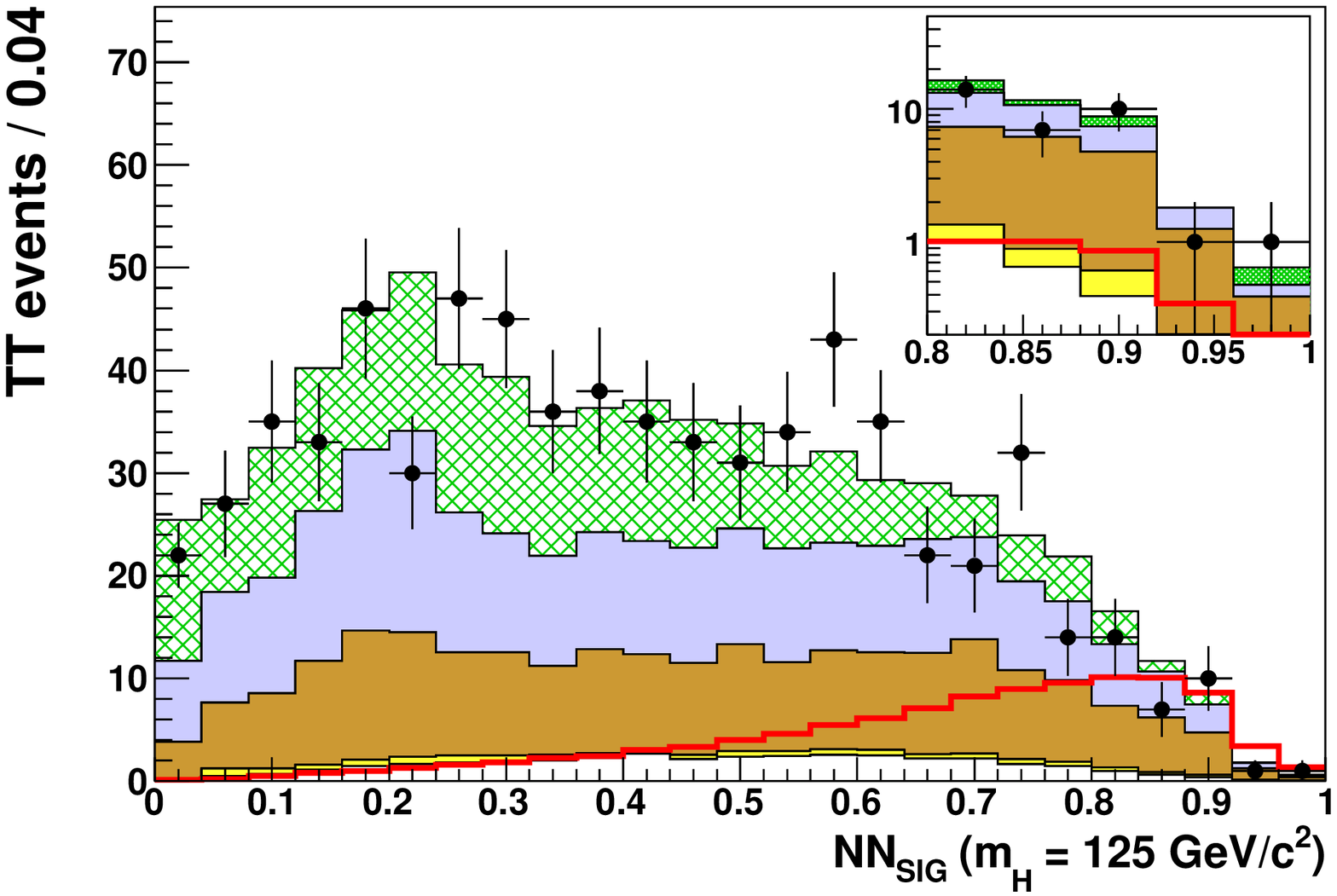}
  \put(210,70){\bf(c)}
  \end{overpic}
}
{
  \begin{overpic}[width=0.55\linewidth]{figures/sigreg_NNSIG_1T_CENT.eps}
  \put(225,80){\bf(a)}
  \end{overpic}
  \begin{overpic}[width=0.55\linewidth]{figures/sigreg_NNSIG_TL_CENT.eps}
  \put(225,80){\bf(b)}
  \end{overpic}
  \begin{overpic}[width=0.55\linewidth]{figures/sigreg_NNSIG_TT_CENT.eps}
  \put(225,80){\bf(c)}
  \end{overpic}
}
\caption{\label{fig:nn_ind} The distributions of tagged data events
and the corresponding expected backgrounds for the \nnsig discriminant
functions after fitting to data for an assumed Higgs boson mass of
125~\gevcc. Panel (a) shows 1T events, (b) shows TL events, and (c)
shows the \nnsig discriminant for TT events.  The signal contribution
(``\vh'') assumes a Higgs boson mass of $125$ \gevcc and is multiplied
by a factor of ten (left unscaled in insets) for illustrative
purposes.  Shown in the inset is a semilogarithmic version of the same
$\nnsig$ distribution for events with $\nnsig > 0.8$.}
\end{figure}

\begin{table*}[t]
  \setlength{\extrarowheight}{3pt}
\begin{ruledtabular} 
\begin{center} 
\caption{\label{tbl:eventYields}
	Comparison of the number of expected and observed events in
	the signal region for different \b-tagging categories.  The
	uncertainties shown include systematic contributions and (when
	appropriate) statistical uncertainties on the simulation
	samples, added in quadrature for a given process. The quoted
	uncertainties for the total expected background prediction
	take into account the appropriate correlations among the
	systematic uncertainties for each background process. Signal
	contributions are given for an assumed Higgs boson mass of 125
	\gevcc.  }
\begin{tabular}{l*{3}{c}}
\toprule
 Process	& { 1T } & { TL } & { TT }  \\
\hline
	QCD multijet 	        &   5941 $\pm$  178 &    637 $\pm$  25  & 222 $\pm$  16  \\ 
	Top 		        &   1174 $\pm$  158 &    302 $\pm$  40  & 271 $\pm$  34  \\ 
	$V$ + heavy flavor jets &   3124 $\pm$  718 &    286 $\pm$  83  & 211 $\pm$  65  \\ 
	Electroweak mistags     &   1070 $\pm$  386 &   \ 55 $\pm$  21  &  13 $\pm$   6  \\ 
        Diboson 	        &    305 $\pm$   46 &     48 $\pm$   6  &  41 $\pm$   5  \\ 
\hline
        Total expected background 
	                        &  11612 $\pm$  949 &   1329 $\pm$ 112  & 759 $\pm$  86  \\ 
\hline
        Observed data 	        &  11955  	    &   1443 		& 692            \\ 
\hline			    
        $\zh\to\vv\bb,\ll\bb$   &     9.7 $\pm$ 1.0 &    5.4 $\pm$ 0.5  & 5.4 $\pm$ 0.5  \\
        $\wh\to\lv\bb$	        &     9.8 $\pm$ 1.0 &    5.3 $\pm$ 0.5  & 5.3 $\pm$ 0.5  \\
\bottomrule
\end{tabular}
\end{center}
\end{ruledtabular}
\end{table*}

\begin{table}[tb]
  \setlength{\extrarowheight}{2pt} 

  \caption{\label{tbl:migrate_signal} Predicted fractions of
  overlapping signal events between the previous analysis and this
  one.  The ``0T/0S'' categories represent events that do not survive
  the tagging or signal-region definition criteria.  Roman-font
  (italicized) numbers represent percentages of overlapping events
  relative to this (the previous) analysis~\cite{metbb}; the sum of
  the percentages in each column (row) is 100\%.  A Higgs boson mass
  of 125~\gevcc is assumed.}

  \begin{tabular*}{\linewidth}{@{\extracolsep{\fill}}cccccccccccc} \hline\hline
               & \multicolumn{2}{c}{0T} & \quad & \multicolumn{2}{c}{1T} & \quad & \multicolumn{2}{c}{TL} & \quad & \multicolumn{2}{c}{TT}   \\ \cline{2-3} \cline{5-6} \cline{8-9} \cline{11-12}
     \emph{0S} \quad &             &     & \quad & ---  & 22\%       & \quad & ---        & 19\%  & \quad & ---  & 6\%  \\
     \emph{1S} \quad & \emph{17\%} & --- & \quad &\emph{63\%} & 67\% & \quad &\emph{15\%} & 31\%  & \quad & \emph{6\% } & 11\% \\ 
     \emph{SJ} \quad & \emph{12\%} & --- & \quad &\emph{20\%} & 9\%  & \quad &\emph{37\%} & 35\%  & \quad & \emph{32\%} & 23\% \\ 
     \emph{SS} \quad & \emph{5\% } & --- & \quad &\emph{3\% } & 1\%  & \quad &\emph{15\%} & 15\%  & \quad & \emph{77\%} & 61\% \\ \hline\hline
  \end{tabular*}
\end{table}

Although the current and previous analyses use the same data set, the
selected event samples used are only partially correlated due to
updates to the \b-tagging algorithm and the \nnqcd discriminant.
Table~\ref{tbl:migrate_signal} shows the predicted fractions of
overlapping signal events between the tag categories of the previous
analysis and those of this one.  As can be seen, only 61\% of the
TT-tagged signal events in this analysis were present in the SS tag
category of the previous analysis.  The remaining 39\% were classified
as SJ events (23\%), 1S events (11\%), or were not analyzed (6\%) due
to either not being tagged or not surviving the minimum
\nnqcd threshold requirement.  A significant portion of TT signal
events is therefore different from the sample of SS events in the
previous analysis.  The percentage of TT data events in this analysis
also present in the SS category of the previous one is approximately
50\%.

The \nnsig discriminant functions trained in the previous
analysis~\cite{metbb} are well modeled in the analogous
\textsc{hobit} categories and also provide good separation of
signal and background events; they were thus retained for this
analysis.  The \nnsig discriminant accepts kinematic and angular
quantities as input variables, as well as the \nnqcd value and a
neural-network output that attempts to disentangle intrinsic \met from
instrumental \met by using tracking
information~\cite{Potamianos:2011}.  The modeling of each input
variable is validated with tagged data in the signal region.
Figure~\ref{fig:nn_ind} shows the \nnsig distribution in the signal
region ($\nnqcd > 0.6$) for the 1T, TL, and TT events after the
discriminants from all tag categories were jointly fitted to data.

\section{Results}

We perform a binned likelihood fit to search for the presence of a
Higgs boson signal.  A combined likelihood is formed from the product
of Poisson probabilities of the event yield in each bin of the
\nnsig distribution for each tag category.  Systematic uncertainties
are treated as nuisance parameters and incorporated into the limit by
assuming Gaussian prior probabilities, centered at the nominal value
of the nuisance parameter, with an RMS width equal to the absolute
value of the uncertainty. The dominant systematic uncertainties arise
from the normalization of the $V$-plus-heavy-flavor background
contributions (30\%), differences in \b-tagging efficiencies between
data and simulation (8--16\%)~\cite{hobit}, uncertainty on the top
(6.5--10\%) and diboson (6\%) cross sections~\cite{top,diboson},
normalizations of the QCD multijet background (3--7\%), luminosity
determination (6\%)~\cite{lumi_cdf}, jet-energy scale
(6\%)~\cite{Bhatti_2005ai}, trigger efficiency (1--3\%), parton
distribution functions (2\%), and lepton vetoes (2\%). Additional
uncertainties applied only to signal include those on the Higgs boson
production cross section (5\%)~\cite{hbb_prod} and on initial- and
final-state radiation effects (2\%).  Also included are uncertainties
in the \nnsig shape, which arise primarily from variations in the
jet-energy scale and the QCD multijet background model.

\begin{figure}[tb]
\includegraphics[width=\linewidth]{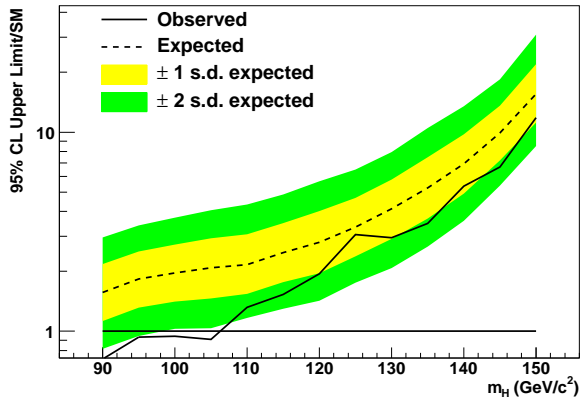}
\caption{\label{fig:limitPlot}Observed and expected (median, for the background-only hypothesis)
95\% C.L. upper limits on $\vh$ cross section times
$\mathcal{B}(H\to b\bar{b})$ divided by the SM prediction, as a
function of the Higgs boson mass. The bands indicate the 68\% and 95\%
credibility regions where the limits can fluctuate, in the absence of
signal.}
\end{figure}

A Bayesian likelihood method is used to set 95\% credibility level
(C.L.) upper limits on the SM Higgs boson production cross section
times branching fraction $\sigma(\vh)\times\mathcal{B}(H\to \bb)$.
For the signal hypothesis, a flat, non-negative prior probability is
assumed for the number of selected Higgs boson events.  The Gaussian
priors of the nuisance parameters are truncated at zero to ensure
non-negative event yield predictions in each \nnsig bin.  The 95\%
C.L. limits for the observed data and the median-expected outcomes
assuming only SM backgrounds are shown in Fig.~\ref{fig:limitPlot} and
Table~\ref{tbl:limits}.  An average improvement of 14\% is obtained in
expected upper limits relative to the previous analysis~\cite{metbb}.
The observed limits lie below the expected values at the level of
roughly one standard deviation for $m_H \ge 120$ \gevcc, and at the
level of approximately two standard deviations for lower Higgs boson
masses.  In constrast, the observed limits of the previous analysis
exceed the median-expected limits by roughly one standard deviation
for $m_H > 120$ \gevcc and are in approximate agreement with expected
limits for lower masses.  These differences correspond to a decrease
of roughly 55\% in the observed limits relative to those of the
previous analysis~\cite{metbb} independent of $m_H$.

\begin{table*}[tb]
  \setlength{\extrarowheight}{3pt}
  \caption{\label{tbl:limits}Expected and observed 95\% C.L. upper limits on the $\vh$ cross section times $\mathcal{B}(H\to b\bar{b})$ divided by the SM prediction~\cite{tevcomb2012}.}
  \begin{ruledtabular}
    \begin{tabular}{cccccccccccccc}
    \toprule
      $m_H$ (\gevcc) & 90 & 95 & 100 & 105 & 110 & 115 & 120 & 125 & 130 & 135 & 140 & 145 & 150 \\ \hline
      Expected & 1.57 & 1.83 & 1.96 & 2.08 & 2.16 & 2.48 & 2.80 & 3.33 & 4.13 & 5.26 & 6.93 & 9.91 & 15.55 \\ 
      Observed & 0.72 & 0.94 & 0.94 & 0.91 & 1.32 & 1.53 & 1.94 & 3.06 & 2.95 & 3.49 & 5.35 & 6.69 & 11.82 \\           
      \bottomrule
    \end{tabular}
  \end{ruledtabular}
\end{table*}

\section{Discussion of Results}

We have investigated potential causes for the sizable shift in the
observed limits.  To quantify the impact of changes to the analysis
design and treatment of systematic uncertainties, we reanalyze the
data sample using the 1S, SJ, and SS categories used in the previous
analysis (Sec.~\ref{sec:redo}).  We also study the effects from other
sources that can influence the observed limits
(Sec.~\ref{sec:crosscheck}).  A summary of the discussion is given in
Sec.~\ref{sec:summary}.

\subsection{\label{sec:redo}Reanalysis using 1S, SJ, and SS tagging categories}

\begin{figure}[tb]
\includegraphics[width=\linewidth]{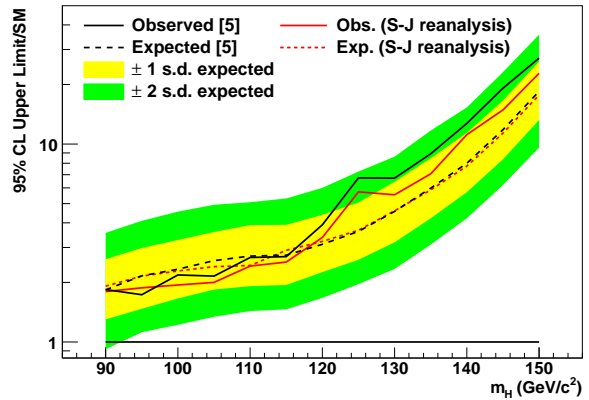}
\caption{\label{fig:limitPlot_redo}Observed and expected (median, 
for the background-only hypothesis) 95\% C.L. upper limits on Higgs
production in the previous analysis~\cite{metbb} and those of the S-J
reanalysis described in Sec.~\ref{sec:redo}. The darker (black) set of
lines represent the observed and expected limits from the previous
analysis, whereas the lighter set (red) represent those of the
S-J reanalysis.  The 68\% and 95\% credibility regions are those of
Ref.~\cite{metbb}.}
\end{figure}

Besides the change in \b-tagging method, there are other less
significant changes made in this analysis with respect to the previous
one:
\begin{enumerate}
\item The \b-tag scale factors and their associated uncertainties 
are now handled with an improved treatment of the correlations between
tag categories.
\item Instead of treating the normalization uncertainties of all 
$V$-plus-heavy-flavor samples as fully correlated, the
$V$-plus-heavy-flavor samples are grouped according to flavor content
of the final state, with each group receiving a 30\% uncertainty.  The
uncertainties associated with each $V$-plus-heavy-flavor group are
treated as uncorrelated with one another.
\item An additional $\met > 35$~\gev requirement is made that corresponds 
to the trigger-level reconstructed \met value.  This has the effect of
further reducing the QCD multijet background at the few percent level.
\item As mentioned in Sec.~\ref{sec:evsel}, upper limits are imposed on 
jet transverse energies. This is done to avoid a kinematic region
susceptible to significant false-positive tagging rates for the
\textsc{hobit} algorithm.
\item An additional $Z$-plus-jets sample is included where the $Z$ boson 
decays to a \bb pair.  The change in overall expected yields due to
this additional sample is very small as the \met here is instrumental.
\end{enumerate}
To estimate the effect of these changes on the limits, we reanalyze
the same data sample using the 1S, SJ, and SS tagging categories of
the previous analysis.  For this test, hereafter referred to as the 
\emph{S-J reanalysis}, we retain the \nnqcd discriminant of the previous 
analysis so that the signal region definitions of this test and that
of the previous analysis are the same.  The results are shown in
Fig.~\ref{fig:limitPlot_redo}.  As can be seen, the expected limits of
Ref.~\cite{metbb} and the S-J reanalysis are in very good agreement.
The observed limits of the S-J reanalysis are systematically lower
than the observed limits of Ref.~\cite{metbb} with an average
difference of $-5\%$ for $m_H < 120$~\gevcc and $-17\%$ for $m_H \ge
120$~\gevcc.
For comparison, we note that the observed limit for the analysis
described in this note is 47\% lower than that of the S-J reanalysis
at $m_H = 125$~\gevcc.  The analysis changes described here thus
account for a non-negligible percentage of the sizable shift in the
observed limits.

We have also investigated the impact of these changes on previously
published combined CDF \hbb limits~\cite{hbb_cdf}.  The \nnsig
discriminants of the S-J reanalysis, and the updated treatment of
systematic uncertainties, are combined with the discriminants of the
CDF $\lv\bb$ and $\ll\bb$ analyses~\cite{lvbb,llbb} to obtain an
updated CDF \hbb result.  Using the discriminants of the S-J
reanalysis, the local significance of the CDF-combined excess at a
Higgs boson mass of 125~\gevcc is recalculated.  Within the
statistical precision of the calculation, the local significance is
unchanged at $2.7$ standard deviations with respect to the
background-only hypothesis.

\subsection{\label{sec:crosscheck}Additional cross-checks}

\subsubsection{\label{sec:btag_effects} Systematic effects from \b-tagging}

Since switching to a new \b-tagging algorithm is the most significant
change adopted for this analysis, it is important to ensure that the
performance of the \textsc{hobit} algorithm is well understood and
well modeled.  As with other \b-tagging algorithms, systematic effects
associated with using \textsc{hobit} are taken into account by
correcting the simulation for differences in \b-tagging behavior
between data and simulation.  Two methods are used to calibrate the
simulation, both of which have been used extensively at CDF: one where
the $\tt$ cross section is fixed to its theoretical prediction, and
scale factors are derived that correct the simulation to the \b-tag
and mistag efficiencies measured in data; and another where heavy- and
light-flavor jets are identified with and without electron conversions
within them, allowing for a determination of the same scale
factors~\cite{hobit}.  As both methods give consistent results for the
\textsc{hobit} scale factors at both T and L operating points, they
are averaged together, resulting in \b-tag efficiency scale factors of
$0.915 \pm 0.035$ (T) and $0.993 \pm 0.035$ (L) and mistag efficiency
scale factors of $1.50 \pm 0.031$ (T) and $1.33 \pm 0.015$ (L), where
the dominant contributions to the uncertainties are from the
theoretical uncertainty on the \tt cross section~\cite{top_hobit}.
The variation of these scale factors with respect to several variables
(e.g., jet energies and instantaneous luminosity) has been
investigated, and any sizable deviations relative to the central
predictions are included in the systematic uncertainties.  These scale
factors and their associated uncertainties have been propagated
through this analysis in a manner consistent with the treatment of
\b-tag and mistag scale factors in the other \hbb CDF
analyses~\cite{llbb,lvbb}.

\begin{figure}
\ifthenelse{\equal{\PrlFigSpacing}{true}}
{
  \begin{overpic}[width=\linewidth]{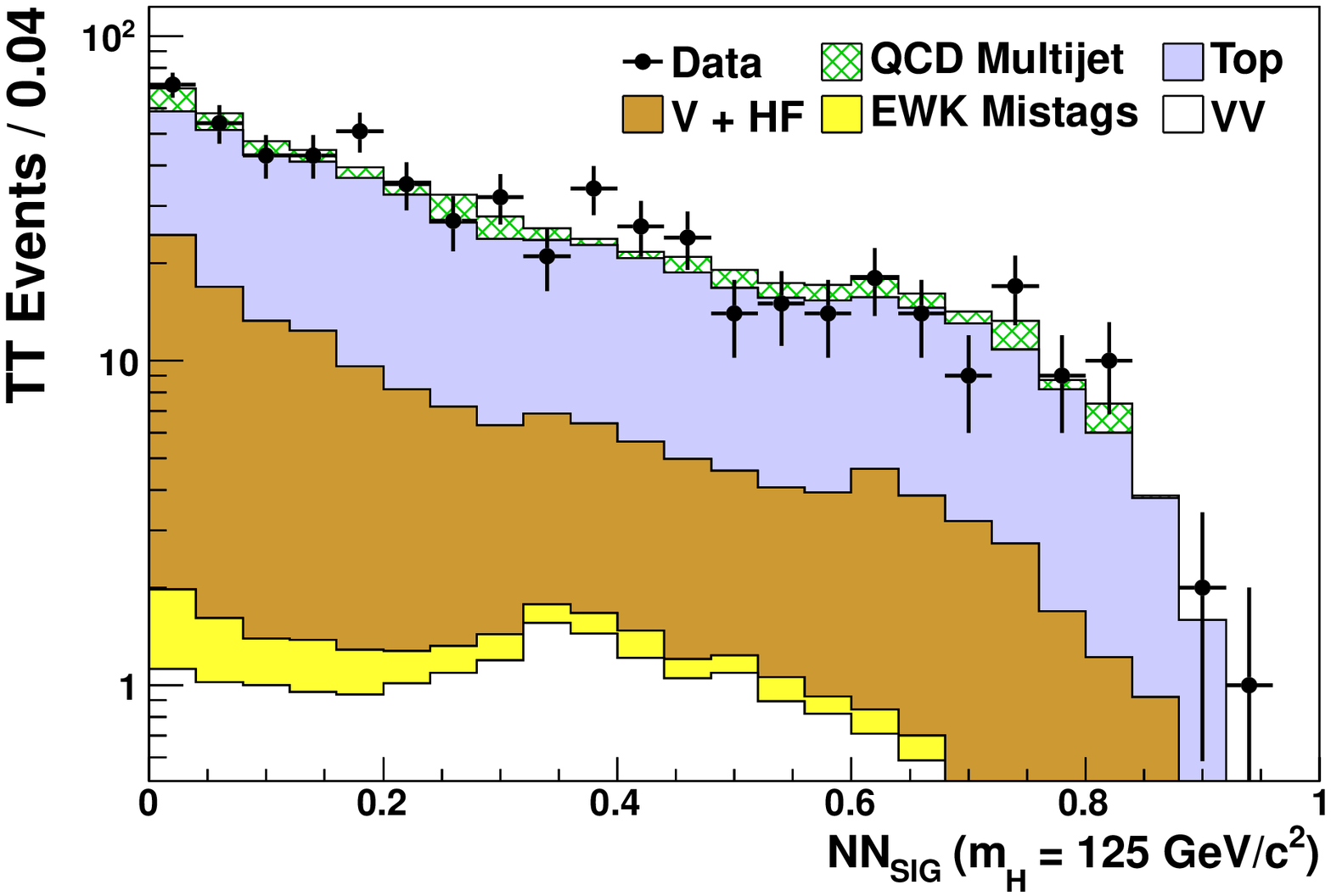}
  \put(217,120){\bf(a)}				     				     
  \end{overpic}					     					     
  \begin{overpic}[width=\linewidth]{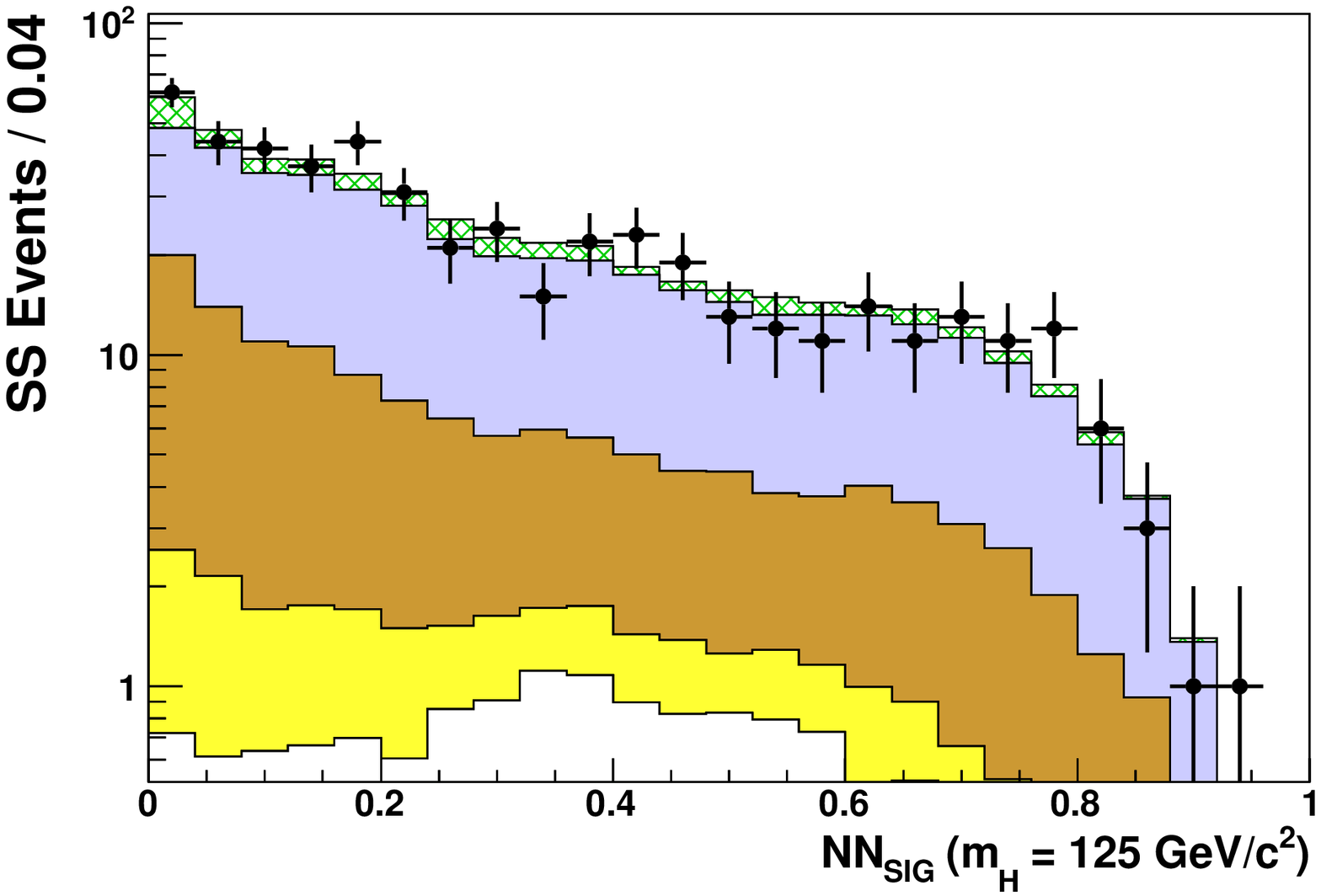}
  \put(217,138){\bf(b)}
  \end{overpic} 
}
{
  \begin{overpic}[width=0.6\linewidth]{figures/ewk_NNSIG_TT_CENT_LOG.eps}
  \put(217,120){\bf(a)}				   					
  \end{overpic}					   						
  \begin{overpic}[width=0.6\linewidth]{figures/ewk_NNSIG_SS_CENT_LOG.eps}
  \put(217,155){\bf(b)}
  \end{overpic} 
}
\caption{ Validation of the background model for (a) TT events and (b) 
reanalyzed SS events in the electroweak control region.}
\label{fig:ewk_nnsig}
\end{figure}

To verify that the choice of \b-tagging algorithm does not result in
mismodeling within the high-score regions of the \nnsig distributions,
we validate the background model with the data in an {\em electroweak
control sample}.  For this control sample we require, in addition to
the preselection sample criteria, the presence of at least one
identified, isolated electron or muon with a minimum transverse
momentum of 20 GeV$/c$ in the event.  The electroweak sample is
dominated by backgrounds that are modeled by simulation and not the
QCD multijet background, whose model is derived from data.
Figure~\ref{fig:ewk_nnsig} shows the \nnsig distributions for TT and
reanalyzed SS events in the electroweak control region.  As can be
seen, there is no obvious difference in the simulation modeling of the
\nnsig discriminants for the \textsc{hobit} or \textsc{secvtx}
algorithms.  Comparisons in the 1T-1S and TL-SJ categories give
similar conclusions.

\subsubsection{\label{sec:stat}Effects of statistical fluctuations}

The expected limits are most significantly impacted by the bins of the
discriminants with the highest signal-to-background ratios.  For the
\nnsig distributions, these are the bins with the highest
\nnsig values, as can be seen in Fig.~\ref{fig:nn_ind}.  Because 
these bins tend to contain only small numbers of data events, the
observed limits are susceptible to statistical fluctuations.  Although
we do not know if the data events are from signal or background
processes, we explore how a fluctuation of yields from either type of
process would manifest itself in the \nnsig distributions.  As part of
the shift in observed limits is due to the analysis changes mentioned
in Sec.~\ref{sec:redo}, the yields quoted below for the SS and SJ
results reflect those of the S-J reanalysis and not those of
Ref.~\cite{metbb}.

As shown in Table~\ref{tbl:migrate_signal}, we expect significant
signal event migrations between the tag categories of the previous
analysis and those of this one.  Consequently, if a Higgs boson signal
is present, we may observe some very high \nnsig score events in one
version of the analysis that either migrate to another tag category or
do not appear within the other analysis.  Since the impact of these
high-score events on the observed limits can be significant, the
migration of a few signal-like events between tag categories in the
S-J reanalysis and the current analysis can lead to non-negligible
changes in observed limits relative to expectations.  Focusing on
discriminant outputs for the 125~\gevcc Higgs boson mass hypothesis,
we compare data events in the very highest-score \nnsig bins of both
analyses and find one potential example for this type of event
migration.  In particular, we observe three events with \nnsig values
above 0.9 in the SJ category that are not present in any tag category
of the current analysis (the new tagging algorithm categorizes two of
these events as LL and the other as 1L).  If these three data events
were to be simply added back into the TL category of the new analysis,
the decrease in the observed limits at $m_H =125$~\gevcc with respect
to those of the S-J reanalysis would be reduced from 47\% to 31\%.

The number of expected background events in the high-score region of
the \nnsig discrimimants is also small and therefore an additional
source of potential statistical fluctuations in the data that might
significantly impact the observed limits.  We check for a potential
effect from background event fluctuations on the difference between
observed limits of the $m_H =$~125~GeV/$c^2$ searches by comparing the
number of observed events that satisfy $\nnsig > 0.8$ to the fitted
background predictions for each tag category in the current analysis
and the S-J reanalysis.  For the most sensitive double-tag categories,
the predicted (observed) event yields in the high-score \nnsig region
are $37.6 \pm 4.6$ (37) for SS and $45.6 \pm 5.1$ (62) for SJ and
$39.5 \pm 4.6$ (33) for TT and $67.4 \pm 6.8$ (80) for TL.  While the
SJ and TL categories exhibit similar upward fluctuations in data
relative to expectations, the data in the SS (TT) category are
consistent with (lower than) the background expectation.

A simple test is performed in which 5~data events are added into the
high-score region of the TT \nnsig distribution (maintaining the
relative fractions of observed events within each high-score bin) to
approximately match the expected background, as was observed in the SS
category.  This change reduces the difference between the present and
S-J reanalyzed limits to 33\%.  Combining this effect with that of
adding the 3 formerly SJ-classified events into the TL category gives
a decrease in observed limits of 19\% relative to the S-J analysis.
This is in reasonable agreement with the expected improvement,
identifying these two effects in data as the primary source of the
change in observed limits at $m_H = 125$~\gevcc.

\begin{table}[tb]
  \setlength{\extrarowheight}{2pt} 

  \caption{\label{tbl:data_corr} Percentages of overlapping events 
  between tag categories of this analysis and the previous one for data
  events with \nnsig values greater than 0.8.}

  \begin{tabular*}{\linewidth}{@{\extracolsep{\fill}}cccc} \hline\hline
        &  1T    &  TL    &  TT   \\ \hline
     1S &  55\%  &  35\%  &  15\% \\ 
     SJ &  \ 4\%  &  20\%  &  30\% \\ 
     SS &  \ 1\%  &  14\%  &  51\% \\ \hline\hline
  \end{tabular*}
\end{table}

\begin{figure}[tb]
\includegraphics[width=\linewidth]{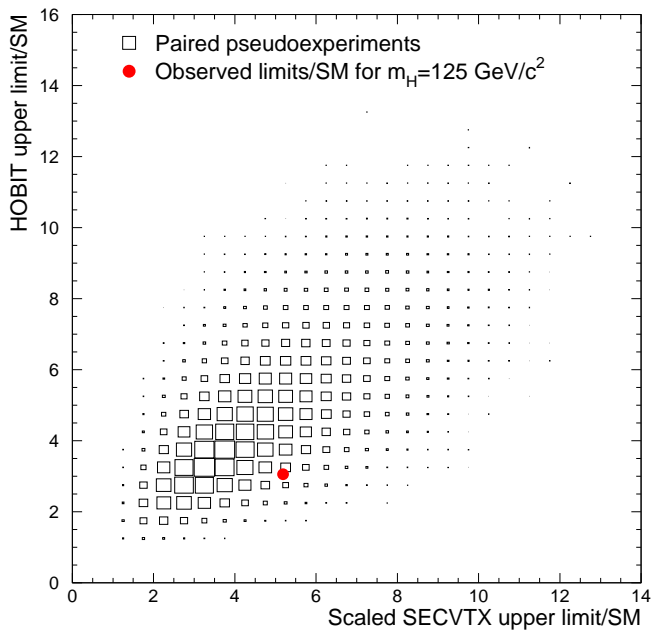}
\caption{\label{fig:pvalue}Pseudoexperiment pairs of expected 95\% C.L. upper limits on Higgs production
assuming the \textsc{hobit} analysis (ordinate) and S-J reanalysis
(abscissa).  For ease in $p$-value computation, the expected limits of
the S-J reanalysis are rescaled such that the median-expected limit
agrees with that of the \textsc{hobit} analysis.  }
\end{figure}

To estimate the probability of an underlying statistical effect
causing such a sizable change in observed limits, correlations between
the event samples must be understood.  For technical reasons we are
not able to determine these correlations separately for each
background process.  Instead, we look directly at the data in the
high-score regions of the \nnsig discriminants, and calculate the
percentage overlap between the tag categories of this analysis and
those of the S-J reanalysis.  The overlap percentages, relative to the
current analysis, are given in Table~\ref{tbl:data_corr}.  Based on
these percentages, we use simulated data experiments to estimate the
probability that the observed limits of this analysis and the S-J
reanalysis are compatible.  Figure~\ref{fig:pvalue} shows a
two-dimensional distribution of expected upper limits, obtained from
producing pairs of expected outcomes between the \textsc{hobit}
analysis and S-J reanalysis.  To calculate a compatibility probability
($p$-value), the probability is estimated for the \textsc{hobit}
analysis to be as or more discrepant that what is observed, given the
observed limit of the S-J reanalysis.  The two-sided probability for
this type of occurrence at a Higgs boson mass of 125~\gevcc is roughly
7\%.

As a downward shift in observed limits is seen across the entire range
of tested $m_H$ values and not just at $m_H = 125$~\gevcc, the
probability for such a global shift to occur must be estimated.
Limited experimental resolution of kinematic event input variables to
the multivariate discriminants leads to events being shared within the
high-score $\nnsig$ regions of the outputs for neighboring mass
hypotheses.  Because of this, we estimate that the number of
independent search regions within our tested Higgs boson mass range
lies somewhere between two and three.  We therefore perform the
pseudoexperiment study for three Higgs boson mass assumptions,
obtaining $p$-values at $m_H = 100$, $125$ and $150$~\gevcc.  Each
$p$-value is on the order of 10\%.  To estimate an approximate global
probability, we combine the obtained $p$-values for the three Higgs
boson mass assumptions using Fisher's method for combining independent
tests.  We obtain a global probability of roughly 3\% or 5\% depending
on whether the number of independent kinematic search regions is three
or two, respectively.

\subsubsection{Background modeling}

In order to conclude that the observed effect in data originates from
statistical fluctuations as opposed to potential background
mismodeling, we confirm the robustness of our background model in
several data control samples.  Events in the intermediate-score region
of the \nnsig distributions are also useful for testing the background
modeling.  We compare predicted and observed event yields in the
\nnsig score region between 0.5 and 0.8, which contains higher event
yields but is above the low-score event region, which drives the
fitted normalizations of the background contributions.  Assuming a
Higgs boson mass of 125~\gevcc, the predicted (observed) event yields
in the intermediate score \nnsig region are $228.8 \pm 21.0$ (217) for
SS and $312.5 \pm 22.6$ (291) for SJ in the S-J reanalysis and $264.8
\pm 25.1$ (265) for TT and $506.1 \pm 38.8$ (506) for TL in the
current one.  Good agreement between the observed and predicted event
yields is found at the other Higgs boson mass assumptions as well.  In
the intermediate-score regions, there is thus no indication of a
background modeling problem that could account for such sizable shifts
in observed limits with respect to the S-J reanalysis.

\subsection{\label{sec:summary}Summary of discussion}

To summarize, the observed limits are very sensitive to statistical
fluctuations in the highest-value bins of the \nnsig distributions.
There is no evidence of any significant mismodeling of the
\textsc{hobit} \b-jet identification algorithm, or of the \nnqcd or \nnsig
distributions and the distributions of their respective input
variables in any of the control regions studied.  The observed
migration of events across the \b-tag categories is fairly consistent
with expectations derived from simulation.  In the most sensitive tag
category, TT, the data yield is about 1 standard deviation below the
background prediction in the signal region.  Using an ensemble of
simulated experiments, we estimate the probability that the observed
limit could change, relative to the S-J reanalysis, by an amount at
least as large as that observed due to statistical fluctuations alone
is about 5\%.  We conclude that the change in the observed limits relative
to the previous analysis is primarily due to statistical fluctuations.

\section{Conclusion}
In conclusion, we have performed an updated Higgs boson search in the
\metbb final state, using the full CDF data set and an improved \b-tagging 
algorithm.  With respect to the previous analysis~\cite{metbb}, the
expected 95\% C.L. limits have improved by 14\% on average across the
Higgs boson mass range $90\le m_H \le 150$ \gevcc.  The 95\% observed
upper limit at a Higgs boson mass of 125~\gevcc is a factor of 3.06
times the SM prediction.  The results of this analysis correspond to
some of the most sensitive limits obtained on Higgs boson production
in the \bb final state.


\section{Acknowledgments}
We thank the Fermilab staff and the technical staffs of the
participating institutions for their vital contributions. This work
was supported by the U.S. Department of Energy and National Science
Foundation; the Italian Istituto Nazionale di Fisica Nucleare; the
Ministry of Education, Culture, Sports, Science and Technology of
Japan; the Natural Sciences and Engineering Research Council of
Canada; the National Science Council of the Republic of China; the
Swiss National Science Foundation; the A.P. Sloan Foundation; the
Bundesministerium f\"ur Bildung und Forschung, Germany; the Korean
World Class University Program, the National Research Foundation of
Korea; the Science and Technology Facilities Council and the Royal
Society, UK; the Russian Foundation for Basic Research; the Ministerio
de Ciencia e Innovaci\'{o}n, and Programa Consolider-Ingenio 2010,
Spain; the Slovak R\&D Agency; the Academy of Finland; the Australian
Research Council (ARC); and the EU community Marie Curie Fellowship
contract 302103.


\bibliography{VH_PRD}

\begin{thebibliography}{28}
\bibitem[{geo()}]{geometry}
\bibinfo{note}{We use a cylindrical coordinate system with the origin at the
  center of the CDF detector, $z$ pointing in the direction of the proton beam,
  $\theta$ and $\phi$ representing the polar and azimuthal angles,
  respectively, and pseudorapidity defined by $\eta= - \rm ln$ $\rm
  tan(\theta/2)$. The transverse momentum $p_T$ (transverse energy $E_{T}$) is
  defined to be $p\sin\theta$ ($E\sin\theta$).}
\bibitem[{MET()}]{MET}
\bibinfo{note}{The calorimeter missing transverse energy $\vmet(\mbox{cal})$ is
  defined by the sum over calorimeter towers, $\vmet(\mbox{cal}) = - \sum_{i}
  E_T^i \hat{n}_i$, where $i$ is a calorimeter tower number with $|\eta| <
  3.6$, $\hat{n}_i$ is a unit vector perpendicular to the beam axis and
  pointing at the $i$th calorimeter tower. The reconstructed missing transverse
  energy, $\vmet$, is derived by subtracting from $\vmet(\mbox{cal})$
  components of the event not registered by the calorimeter, such as jet energy
  adjustments. $\met$ [$\met(\mbox{cal})$] is the scalar magnitude of $\vmet$
  [$\vmet(\mbox{cal})$].}
\bibitem[{crs()}]{crs}
\bibinfo{note}{We validate the background model by comparing with tagged data
  events in the preselection sample, the electroweak control sample
  (Sec.~\ref{sec:btag_effects}) and also in the MJ1, MJ2, and MJ3 control
  regions as defined in Ref.~\cite{metbb}.}
\bibitem[{sph()}]{sphericity}
\bibinfo{note}{The jet sphericity is defined as $S = \frac{3}{2}(\lambda_2 +
  \lambda_3)$, where $\lambda_{2,3}$ are eigenvalues of the sphericity tensor
  $S^{\alpha\beta} = \sum_i p_i^\alpha p_i^\beta/\sum_i p_i^2$, with the sums
  extending over each jet $i$ with transverse energy greater than 15~\gev. The
  eigenvalues are ranked according to $\lambda_1 > \lambda_2 > \lambda_3$, and
  they satisfy $\lambda_1 + \lambda_2 + \lambda_3 = 1$.}
\end{thebibliography}

\end{document}